\documentclass[aps,prb,twocolumn,superscriptaddress,showpacs,floatfix]{revtex4-1}
\usepackage[english]{babel}
\usepackage{graphicx}
\usepackage{stmaryrd}
\usepackage{amssymb}
\usepackage{amsfonts}
\usepackage{amsmath}
\usepackage{color}
\usepackage{xspace}

\usepackage{txfonts}
\usepackage{bm}

\DeclareSymbolFont{largesymbols}{OMX}{cmex}{m}{n}

\DeclareGraphicsExtensions{.pdf,.ps,.eps}

\newcommand{\calS}{{\mathcal{S}}}

\begin{document}

\title{Single-ion anisotropy in Haldane chains and
form factor of the O(3) nonlinear sigma model}

\author{Shunsuke C. Furuya}
\affiliation{Institute for Solid State Physics, University of Tokyo,
Kashiwa 277-8581, Japan}
\author{Takafumi Suzuki}
\affiliation{Research Center for Nano-Micro Structure Science and
Engineering, Graduate School of Engineering, University of Hyogo, Himeji
671-2280, Japan}
\author{Shintaro Takayoshi}
\affiliation{Institute for Solid State Physics, University of Tokyo,
Kashiwa 277-8581, Japan}
\author{Yoshitaka Maeda}
\affiliation{Analysis Technology Center, Fujifilm Corporation,
Kanagawa 250-0193, Japan}
\author{Masaki Oshikawa}
\affiliation{Institute for Solid State Physics, University of Tokyo,
Kashiwa 277-8581, Japan}

\date{\today}
\begin{abstract}
We consider spin-1 Haldane chains with single-ion anisotropy,
which exists in known Haldane chain materials.
We develop a perturbation theory in terms of anisotropy,
where the magnon-magnon interaction is important even in the
low temperature limit.
The exact two-particle form factor in
the O(3) nonlinear sigma model leads to quantitative
predictions on several dynamical properties, including the
dynamical structure factor
and electron spin resonance frequency shift.
These agree very well with numerical results, and with experimental
data on the Haldane chain material Ni(C$_5$H$_{14}$N$_2$)$_2$N$_3$(PF$_6$).
\end{abstract}

\maketitle


One-dimensional quantum spin systems are an ideal subject
to test sophisticated theoretical concepts against experimental
reality.~\cite{giamarchi_book_1d}
One of the best examples is the Haldane gap problem.
Haldane predicted in 1983 (Ref.~\onlinecite{Haldane-conj})
that the standard Heisenberg antiferromagnetic (HAF) chain
$ \mathcal H = J \sum_j \bm S_j \cdot \bm S_{j+1}$
has a non-zero excitation gap and exponentially decaying
spin-spin correlation function for an integer spin quantum
number $S$.
It has been long known that the HAF chain
with $S=1/2$ is exactly solvable by a Bethe ansatz, and that it
has gapless excitations and the power-law
spin-spin correlation function.
While the same model cannot be solved exactly
for $S \geq 1$, Haldane's prediction was
rather unexpected and surprising at the time.

Haldane's argument was based on the mapping of the HAF
chain to the O(3) nonlinear sigma model (NLSM), which
is a field theory defined by the action
\begin{equation}
 \mathcal A_0 = \frac 1{2g} \int dt dx\, \biggl[\frac 1{v}(\partial_t
  \bm n)^2 - v (\partial_x \bm n)^2\biggr]
  + i \theta Q,
  \label{eq.action_O3}
\end{equation}
where $g= 2/S$ is coupling constant, $v$ is spin-wave
velocity, $\theta = 2\pi S$ 
and $Q = (1/4\pi)\int dtdx\, \bm n \cdot \partial_t \bm n \times
\partial_x \bm n$ is an integer-valued topological charge.
The field $\bm n(x)$ is related to the spin $\bm S_j$ via $\bm S_j
\approx (-1)^j \sqrt{S(S+1)}\, \bm n(x) + \bm L(x)$, where
$\bm L(x) = \bm n \times \partial_t \bm n/g$.
The field $\bm n$ has a constraint $\bm n^2 = 1$.
For a half-integer $S$, the topological term $i\theta Q$ should be kept.
However, for an integer $S$, the topological term 
$i\theta Q = 2\pi i \times (\text{integer})$ is
irrelevant and it suffices to drop $i \theta Q$ in
eq.~\eqref{eq.action_O3}.
The O(3) NLSM without the topological term  is 
a massive field theory, which implies that the integer $S$ HAF chain 
(Haldane chain) has
a non-zero gap and a finite correlation length.
The Haldane's conjecture is now confirmed by a large body of
theoretical, numerical, and experimental
studies.~\cite{Affleck-Haldanegap-review}
Moreover, the O(3) NLSM is also useful in describing integer $S$ HAF
chains.

There are various complications in real materials.
A Haldane chain material generally has a single-ion anisotropy (SIA):
$
\mathcal H' = \sum_j [ D(S^z_j)^2 + E \{ (S^x_j)^2 - (S^y_j)^2\}]
$.
This interaction is important,
for example, for electron spin resonance (ESR)
measurements.
ESR is a useful experimental probe
which can detect even very small anisotropies.
In other words, the anisotropic interaction is the key
to understanding a rich store of ESR experimental data.
However, the theory of ESR is not sufficiently developed
for many systems, including Haldane chains,
leaving many experimental data not being understood.
In order to fully exploit the potential of ESR,
accurate formulation of the SIA in Haldane chains is required.

The SIA can be treated as a perturbation since it is usually small
compared to the isotropic exchange interaction $J$.
In the O(3) NLSM language, the perturbation is written as
\begin{align}
 \mathcal H' &=S(S+1) \int dx \bigl[D(n^z)^2 - E \{
 (n^x)^2 - (n^y)^2\}\bigr],
 \label{eq.sia}
\end{align}
which spoils the integrability of the O(3) NLSM.
Several simple calculations have been done
based on the Landau-Ginzburg (LG) model.~\cite{affleck1991,affleck1992esr}
When the elementary excited particles (magnons) 
are dilute, the interaction between magnons
may be ignored. If this is the case, the system is effectively
described by a much simpler theory of free massive magnons
(the LG model).~\cite{affleck1991}
However, description by the LG model is not accurate and, furthermore,
it is phenomenological.~\cite{essler2004haldane}
Even in the low-energy limit, where the free magnon approximation
is supposed to be exact, it is not the case
with respect to the evaluation of Eq.~\eqref{eq.sia}.
This is because the perturbation~\eqref{eq.sia} creates and
annihilates two magnons at the same point;
in such a situation, interaction among the magnons is
indeed important even when the average density of magnons
in the entire system is infinitesimal.
Therefore, correct handling of the SIA in the O(3) NLSM framework
requires a proper inclusion of the magnon interaction.

In this paper, we present such a formulation,
utilizing the integrability of the O(3) NLSM.
The effects of interaction are encoded in the form factors
of operators. 
The form factors in integrable field theories
can be determined by the consistency with the exact $S$-matrices
and several additional
axioms.~\cite{KarowskiWeisz-NPB1978,BergKarowskiWeisz-PRD1979,Smirnov-FF-book}
Form factor expansion (FFE) is particularly powerful in massive
field theories such as the O(3) NLSM, because
the higher-order contributions survive only above
the higher energy thresholds.~\cite{EsslerKonik-FF-massive}
The leading contribution to the FFE
of Eq.~\eqref{eq.sia} is given by the two-particle form factor.
The FFE shows an excellent agreement with
the correlation function of $(S^z)^2$
numerically obtained in the $S=1$ HAF chain,
demonstrating the importance of the interaction.
At the same time, the renormalization factor for the SIA~\eqref{eq.sia}
is determined by the fitting of the numerical data.
Furthermore, we discuss two applications to physical problems of
interest: the split of triplet magnons in the
dynamical structure factor and
the ESR shift in the $S=1$ HAF chain with SIA. 
We find very good agreement with numerical results in both applications,
and with experimental data on the ESR shift, without introducing
any extra fitting parameter.

A single magnon excitation can be parametrized by the rapidity $\theta$,
so that its energy and wavenumber are given
respectively as $\Delta_0 \cosh \theta$ and 
$(\Delta_0/v) \sinh \theta$, where
$\Delta_0 = 0.41J$ is the Haldane gap.
Because of interactions among magnons,the $S$ matrix of O(3) NLSM has a
complicated structure.~\cite{Zamolodchikov1978Smatrix}
The one-particle form factor of an operator $\mathcal O$ is defined as
a matrix element which connects the ground state $|0\rangle$
to a one-particle state $|\theta_1, a_1\rangle$ ($a_1=1,2,3$), namely
$F_{\mathcal O}(\theta_1, a_1) \equiv \langle 0 |\mathcal O|\theta_1, a_1\rangle$.
And the $n$-particle form factor is defined as
$F_{\mathcal O} (\theta_1, a_1; \theta_2, a_2; \cdots; \theta_n, a_n) \equiv
\langle 0 |\mathcal O|\theta_1, a_1; \theta_2, a_2; \cdots;
\theta_n, a_n\rangle$, where this $n$-particle state is normalized as
$\langle \theta'_1, a'_1;\cdots;  \theta'_n, a'_n|\theta_1,
a_1; \cdots; \theta_n, a_n\rangle =
(4\pi)^n \delta_{a'_1, a_1} \cdots \delta_{a'_n, a_n}
\delta(\theta'_1 -\theta_1)  \cdots
\delta(\theta'_n - \theta_n)$. 

The FFE of the fundamental field $n^a$, which
corresponds to (a staggered part of) the spin operator $S^a$,
has often been studied.
The leading contribution to the FFE
is the one-particle form factor $F_{n^a}(\theta_1, a_1)$.
Because $n^a$ is odd under the transformation
$\bm{n} \rightarrow - \bm{n}$, the next order contribution
comes from the three-particle form factor, which gives
small corrections to the spin-spin correlation
function.~\cite{HortonAffleck,essler2000three}
On the other hand, the composite operator $(S^a)^2$,
which is of our central interest, has been less studied.
Since it is proportional to $(n^a)^2$ and
even under the reversal $\bm{n} \rightarrow - \bm{n}$,
the leading contribution to the FFE
comes from the two-particle form factor
$F_{(n^a)^2}(\theta_1, a_1;\theta_2, a_2)$.
We note that the exact two-particle form factor of the antisymmetric
field $\bm L(x)$ in the O(3) NLSM has been applied to describe the
uniform part of the spin-spin correlation function of HAF
chains.~\cite{whiteaffleck,sorensen1994sofk,sorensen1994equaltime,affleck_weston1992,konik2003prb}
Including the renormalization factors for spin operators,
which are undetermined at this point, we have
\begin{align}
 F_{S^a} (\theta_1, a_1) &= \sqrt{Z} \, \delta_{a,a_1} ,
 \label{eq.ff_n} \\
 F_{(S^a)^2} (\theta_1, a_1; \theta_2, a_2)
 &= -i Z_2 \,\delta_{a_1,a_2} ( 3 \delta_{a, a_1} - 
 1)\psi_2(\theta_1 - \theta_2) .
 \label{eq.ff_n2}
\end{align}
The two-particle form factor~\eqref{eq.ff_n2}
receives contributions from higher-order terms
in the FFE of $S^a$, and
cannot be determined by Eq.~\eqref{eq.ff_n} alone.
Thus $Z_2$ is a parameter independent of $Z$.

We have the constraint $\sum_{a=1,2,3} (S^a)^2 = 2$ on the composite operator.
From this constraint and the O(3) symmetry, it follows that
$\sum_{a=1,2,3} F_{(S^a)^2}(\theta_1, 3; \theta_2, 3) = \langle 0|\theta_1, 3;
\theta_2, 3\rangle = 0$, which is satisfied by \eqref{eq.ff_n2}.
Integral representation of $\psi_2(\theta)$ is given 
in Ref.~\onlinecite{balog_weisz2007NPB}, for O($N$) NLSM with a general integer $N$.
For $N=3$, it reads
\begin{equation*}
 \psi_2(\theta) = \sinh \frac\theta 2 \, \exp\left[
  \int_0^\infty \frac{d\omega}\omega e^{-\pi \omega}\frac{\cosh[(\pi +
  i\theta)\omega]-1}{\sinh(\pi \omega)}\right].
\end{equation*}
This integral can be analytically carried out to give
\begin{equation}
 \psi_2(\theta) = \frac i2 (\theta - \pi i) \tanh \frac \theta 2.
  \label{eq.psi2}
\end{equation}

\begin{figure}
 \centering
 \includegraphics[width=\linewidth]{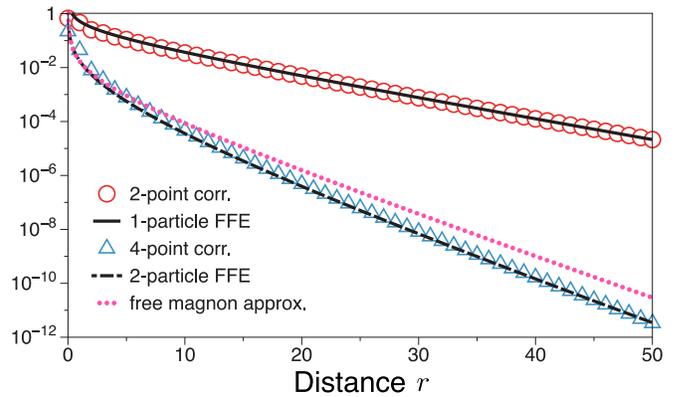}
 \caption{(color online): Numerically calculated
 spin-spin correlation $(-1)^r\langle0| S^z(r) S^z(0)|0 \rangle$
 (circles) and the correlation 
 $\langle 0|(S^z(r))^2 (S^z(0))^2|0\rangle - 4/9$ (triangles)
 are compared with FFEs \eqref{eq.corr2p}
 with $Z=1.26$ (solid curve)
 and the connected part of
 \eqref{eq.corr4p} with $Z_2 = 0.24$ (dashed curve).
 The free magnon approximation (dotted curve) cannot fit
 the correlation function of $(S^z)^2$.
 }
 \label{fig.RGfactors}
\end{figure}

Determination of the renormalization factors $Z$ and $Z_2$ requires
numerical calculations.
In order to test the validity of the FFE for $(S^a)^2$ and
further to determine $Z_2$, 
we computed the equal-time correlation function
$\langle 0|(S^z(r))^2 (S^z(0))^2|0\rangle$
by the infinite time-evolving block decimation (iTEBD)
method,~\cite{vidal2007iTEBD} as shown in Fig.~\ref{fig.RGfactors}.

FFE is derived by
inserting the identity $\hat 1 = \sum_{n=0}^\infty P_n$,
where the $P_n$'s are the projection operators to the $n$-particle subspace of the
Fock space, defined by
$P_0 = |0\rangle \langle 0 |$
and $P_n = \frac 1{n!} \sum_{a_1, \cdots, a_n} \int \frac
{\prod_j d\theta_j}{(4\pi)^n} |\theta_1, a_1; \cdots; \theta_n, a_n\rangle
\langle \theta_1, a_1; \cdots; \theta_n, a_n|$ for $n \ge 1$.
In the leading nonvanishing order, we find
\begin{align}
 &(-1)^r \langle 0 |S^z(r)S^z(0)|0\rangle
 \approx Z \int \frac{d\theta}{4\pi}\,  e^{i\Delta_0 r \sinh \theta/v},
 \label{eq.corr2p} \\
 &\langle 0 |(S^z(r))^2 (S^z(0))^2 |0\rangle - \frac{4}{9} \notag \\
 &\approx 3{Z_2}^2 \int \frac{d\theta_1 d\theta_2}{(4\pi)^2}
 |\psi_2(\theta_1 - \theta_2)|^2 e^{i\Delta_0 r (\sinh \theta_1 + \sinh
 \theta_2)/v}.
 \label{eq.corr4p}
\end{align}
$Z = 1.26$ was given in Ref.~\cite{sorensen1994equaltime}
by comparing numerically obtained spin-spin correlation function with
the LG model.
Concerning the spin-spin correlation function,
the LG model is equivalent to the lowest-order FFE~\eqref{eq.corr2p};
our iTEBD calculation also reproduces the result
of Ref.~\onlinecite{sorensen1994equaltime}. 
On the other hand, to the best of our knowledge, 
$Z_2$ has not been determined previously.

As shown in Fig.~\ref{fig.RGfactors},
the lowest order of FFE~\eqref{eq.corr4p}
shows an excellent
agreement with the numerical data; the fit also determines
\begin{equation}
 Z_2 = 0.24.
  \label{eq.Z2}
\end{equation}
Since we used the known values of the Haldane gap $\Delta_0=0.41J$ and
the spin-wave velocity $v=2.49J$ (Ref.~\onlinecite{todo2001prl}) for $S=1$,
the renormalization factor $Z_2$ is the only fitting parameter.

In contrast to the FFE~\eqref{eq.corr4p},
the LG model, which ignores interaction among magnons,
shows discrepancy with the numerical data, as also shown
in Fig.~\ref{fig.RGfactors}.
To illustrate the effect of the interaction,
let us discuss the asymptotic long-distance behavior
of Eqs.~\eqref{eq.corr2p} and \eqref{eq.corr4p}. 
When $r \to + \infty$, only the behavior of $\psi_2(\theta)$
at $\theta \sim 0$ is relevant in \eqref{eq.corr4p}.
Here we can expand \eqref{eq.corr2p} and \eqref{eq.corr4p} as
$(-1)^r \langle 0 |S^z(r) S^z(0)|0 \rangle \propto
e^{-r/\xi}/\sqrt{8\pi r/\xi}$
and $\langle 0 |(S^z(r))^2 (S^z(0))^2|0\rangle - 4/9 \propto
 e^{-r/\xi_2}/(4\pi r/\xi_2)$.
In a relativistic field theory, the
inverse correlation length is equivalent to the
lowest excitation energy created by the operator;
in fact $\xi = v/\Delta_0$.
Furthermore, in the LG model,~\cite{affleck1991}
$\xi_2 = \xi/2$ should hold.
This is because
the composite field $(n^a)^2$ creates two particles,
and O(3) NLSM does not contain any bound
states.~\cite{bergknoff1979thirring}
Thus the excitation energy for the two-particle creation
would be twice the magnon mass ($2\Delta_0$),
implying $\xi_2=\xi/2$.
However, the actual numerical data are inconsistent with
this relation: $\xi_2 = 2.75 < \xi/2 = 3.01$.
This discrepancy is attributed to the interaction between
magnons. Since $(n^a)^2$ creates two magnons \textit{at the same point},
the actual excitation energy is larger than $2 \Delta_0$,
resulting in $\xi_2 < \xi/2$.

\begin{figure}
 \centering
 \includegraphics[width=\linewidth]{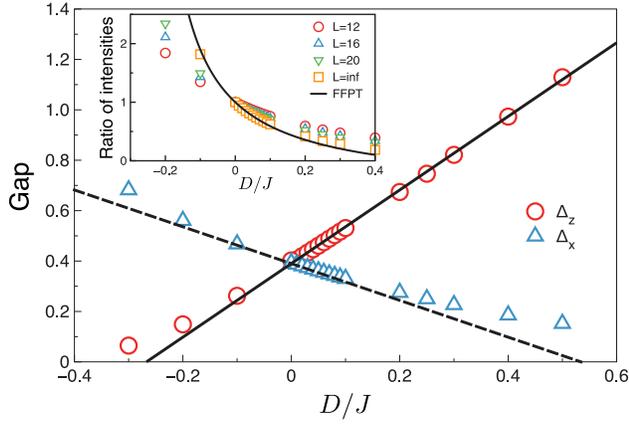}
 \caption{(Color online): Numerically determined excitation gaps
 $\Delta_x$ (circles) and $\Delta_z$ (triangles)
 are plotted for $-0.4 \le D/J \le 0.6$ and $E=0$.
 Deviation ofthe  numerical data from the first order FFPT
  (solid and dashed lines) is attributed
 to higher-order perturbations.
 Inset: The ratio $\mathcal S^{zz}(\pi, \Delta_z)/\mathcal S^{xx}(\pi, \Delta_x)$ 
 obtained by the Lanczos method (symbols) and \eqref{eq.ratio} (solid curve)
 are compared. The extrapolation to $L=\infty$ is
 done by fitting the finite-size
 data for $L=12,14, 16, 18$ and $20$ with a polynomial of $1/L$.
 }
\label{fig:magnonmass}
\end{figure}

With the full determination of the two-particle form factor \eqref{eq.ff_n2},
we turn to discussion of  dynamical structure factor (DSF)
at $T=0$ in Haldane chains with a SIA.
The peaks in the DSF reflect the energy of the magnon
at a given momentum.
Triply degenerate magnon dispersions in
the isotropic chain are split due to the SIA.
We determine the
first-order perturbation to the masses,
$\Delta_a^{(1)} \equiv \Delta_a - \Delta_0$,
in the form-factor perturbation theory (FFPT)~\cite{controzzi2004ffpt}:
\begin{equation}
 \Delta_a^{(1)} \sim
  \frac{\langle \theta, a|\mathcal H'|0, a\rangle}
  {\langle \theta, a|0, a\rangle}.
  \label{eq.delta_a}
\end{equation}
In fact, both the numerator and the denominator are
proportional to $\delta(\theta)$, and Eq.~\eqref{eq.delta_a}
should be understood as the ratio of the coefficients of $\delta(\theta)$.
Furthermore, the numerator equals
to $F_{\mathcal H'}(0, a; \theta -\pi i, a)$
because of the crossing symmetry.~\cite{LeClair1999}
Therefore, \eqref{eq.delta_a} reads
\begin{align}
 \Delta_x^{(1)} &= - \frac{Z_2v}{2\Delta_0}D - \frac{3Z_2v}{2\Delta_0}E,
 \label{eq.deltax_1} \\
 \Delta_y^{(1)} &= - \frac{Z_2v}{2\Delta_0}D + \frac{3Z_2v}{2\Delta_0}E,
 \label{eq.deltay_1}\\
 \Delta_z^{(1)} &= \frac{Z_2v}{\Delta_0}D.
 \label{eq.deltaz_1}
\end{align}

\begin{figure}
 \centering
 \includegraphics[width=\linewidth]{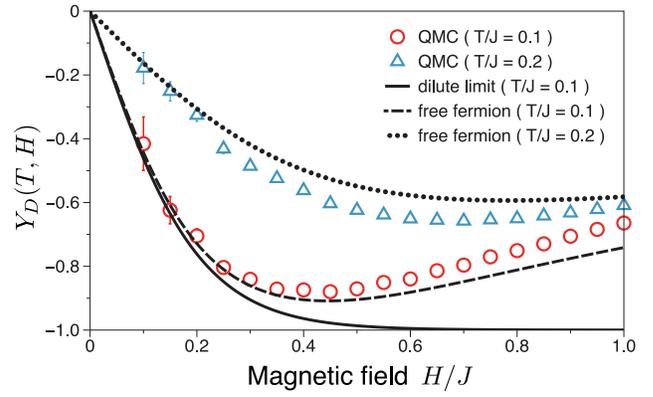}
 \caption{(Color online): Magnetic field dependence of ESR shift
 $Y_D(T,H)$ for $T=0.1J$ (circles) and $T=0.2J$ (triangles).
 The solid curve is \eqref{eq.YD_dilute}, which is exact in $H \to 0$.
 The dashed and dotted curves are \eqref{eq.YD_fermion} at $T=0.1J$ and
 $T=0.2J$, respectively.
 }
 \label{fig.esr}
\end{figure}

The leading contribution to the $T=0$ DSF $\calS^{aa}(\pi, \omega)$
corresponds to the creation of a single magnon.
Therefore we find
\begin{equation}
  \calS^{aa}(\pi, \omega)
  \sim \frac{\pi Zv}{\Delta_a}\delta(\omega - \Delta_a),
  \label{eq.Saa}
\end{equation}
which has the identical form to the DSF of a system of
free particles.
This is natural because the population of magnons approaches
zero in the $T \to 0$ limit, and thus the interactions are
negligible.
Nevertheless, we emphasize that the change of the masses as
$\Delta_a$~\eqref{eq.deltax_1}--\eqref{eq.deltaz_1} due to the SIA
is affected by the magnon-magnon interaction.
Equation~\eqref{eq.Saa} implies that the magnon masses $\Delta_a$
can be identified with the peak frequency of DSF at
the antiferromagnetic wavevector $q=\pi$.
In Fig.~\ref{fig:magnonmass}, we compare the
magnon masses $\Delta_a$ extracted from the $T=0$ DSF peak
obtained numerically by the Lanczos method~\cite{suzuki2005}
for various values of $D$ (while setting $E=0$).
For small $D$, the numerical data agree very well with
the FFPT~\eqref{eq.deltax_1}--\eqref{eq.deltaz_1}.

The form of the $T=0$ DSF~\eqref{eq.Saa} leads to
another prediction: The ratio of the DSF intensities
should obey
\begin{equation}
\frac{\int d\omega \; \calS^{zz}(\pi, \Delta_z) }{
\int d\omega \; \calS^{xx}(\pi, \Delta_x)} = \frac{\Delta_x}{\Delta_z}.
\label{eq.ratio}
\end{equation}
This is also confirmed by the Lanczos data as shown
in the inset of Fig.~\ref{fig:magnonmass}.

Let us extend our discussion to the system under
a finite magnetic field.
Now our Hamiltonian $\mathcal H = \mathcal H_0 +
\mathcal H_Z + \mathcal H'$ consists of three terms.
$\mathcal H_0$ is the SU(2) symmetric exchange interaction,
$\mathcal H_Z = -g_e\mu_B \bm H \cdot \bm S
= - g_e \mu_B\bm H \cdot \sum_j \bm S_j$ is the
 Zeeman interaction,
and $\mathcal H'$ is the SIA, which is assumed to be small.
$g_e$ is Land\'e $g$ factor of electrons 
and $\mu_B$ is the Bohr magneton.
We set $g_e \mu_B = 1$ unless otherwise stated.
ESR is a very powerful tool to study the effects of
anisotropies on spin dynamics.
One of the fundamental quantities in ESR is the
resonance frequency shift (ESR shift).
The ESR shift is generally given, in the first order of the
anisotropy $\mathcal H'$, as~\cite{kanamori1962,nagata1072,maeda2005perturbation}
\begin{equation}
 \delta \omega = - \frac{\langle [[\mathcal H', S^+], S^-]\rangle_0}{2\langle
  S^z\rangle_0}.
  \label{eq.KT}
\end{equation}
$\langle \cdots \rangle_0$ denotes the average with respect to 
the unperturbed Hamiltonian $\mathcal H^{(0)} = \mathcal H_0 + \mathcal H_Z$.
For the SIA, \eqref{eq.KT} reads
$\delta \omega = f(\Theta, \Phi)\, Y_D(T,H)$, where
$f(\Theta, \Phi) = D(1-3\cos^2\Theta) -3E \sin^2\Theta \cos 2\Phi$ and
\begin{equation}
 Y_D(T,H) = \frac{\sum_j \langle 3(S^z_j)^2 - 2\rangle_0}{2\langle
  S^z\rangle_0}.
  \label{eq.KT_SIA}
\end{equation}
$(\Theta, \Phi)$ is the polar coordinate of the magnetic field axis.

\begin{figure}[!t]
 \centering
 \includegraphics[width=\linewidth]{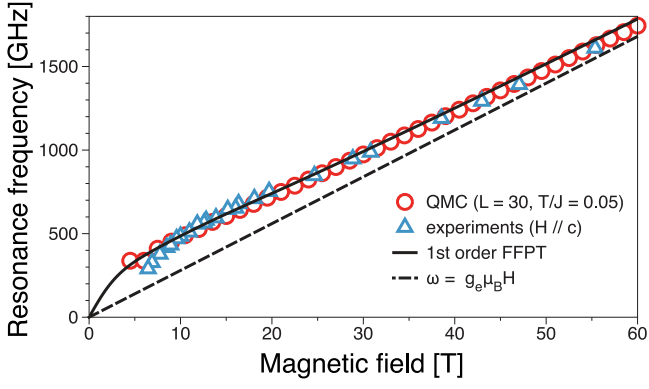}
 \caption{(Color online) Comparison of the resonance frequency
 $\omega_{\mathrm r} = g_e \mu_B H + \delta \omega$ by QMC (circles)
 with experimental
 data~\cite{kashiwagi2009ndmap} (triangles).
 We performed QMC calculations with $L=30$ sites.
 We used $D=0.25J$ and $H \parallel c$ ($\Theta = \Phi = 0$).
 The solid curve is obtained from \eqref{eq.YD_fermion} and
 the dashed line represents the paramagnetic resonance $\omega = g_e
 \mu_B H$.
 }
 \label{fig.ndmap}
\end{figure}

To apply the results of the FFPT,
first we consider the limit $T, H \ll \Delta_0$.
Here we could project the numerator to one-magnon subspace,
ignoring the multi-magnon contributions.
The projection operator is $P_1 = \int \frac{d\theta}{4\pi} \sum_{a=0, \pm}
|\theta, a\rangle \langle \theta, a|$.
Note that we introduce a different set of indices $a=0,\pm$ representing 
magnons with dispersion $E_{a}(\theta)=\Delta_{0}\cosh\theta-aH$.
The projection leads to
\begin{align}
 &P_1 \sum_j \bigl[ 3(S^z_j)^2 - 2\bigr]P_1 \notag \\
 &= \int \frac{d\theta}{4\pi} \frac{3 Z_2 v}{2\Delta_0 \cosh \theta}
 \Bigl[ 2|\theta, 0\rangle \langle \theta, 0| - |\theta, + \rangle
 \langle \theta, +| - |\theta, - \rangle \langle \theta, - |\Bigr].
 \label{eq.projection}
\end{align}
Its thermal expectation value can be given in terms of
the (classical) distribution function.
Thus we find
\begin{align}
 Y_D(T,H) &= -\frac{3Z_2}{4} \tanh \biggl( \frac H{2T}\biggr)
 \frac{\int \frac{d\theta}{4\pi} \frac v{\Delta_0 \cosh \theta}
 e^{-\Delta_0 \cosh \theta/T}}{\int \frac{d\theta}{4\pi} e^{-\Delta_0
 \cosh \theta/T}}.
 \label{eq.YD_dilute}
\end{align}

Figure~\ref{fig.esr} shows the magnetic field dependence of $Y_D(T,H)$,
comparing \eqref{eq.YD_dilute} from the FFPT
with the numerical results obtained by
\eqref{eq.KT_SIA} with quantum Monte Carlo (QMC) method in
ALPS software.~\cite{ALPS-1.3}

Although the agreement is good at low temperature $T=0.1J$
and at low magnetic fields $H \ll \Delta_0$, the discrepancy
is evident for $H \gtrsim \Delta_0$.
This is rather natural, because the magnon population increases
as $H$ is increased, invalidating the dilute limit
approximation made in the derivation of Eq.~\eqref{eq.YD_dilute}.
In particular, $T=0$, $H= \Delta_0$ is a quantum critical point which
separates the low field gapped phase and the high field
TLL phase, where magnons are condensed.
Although it is difficult to handle the case with nondilute magnons, 
a reasonable improvement would be incorporating magnon-magnon repulsion 
through the Pauli exclusion principle by utilizing
the Fermi-Dirac distribution function $f_{a}(k)=[e^{\omega_{a}(k)/T}+1]^{-1}$
instead of the classical one, in Eq.~\eqref{eq.YD_dilute}.
This is demonstrated by the fact that the $z=2$ 
free-fermion theory well describes the low-energy behavior
near the quantum critical point
$H=\Delta_{0}$.~\cite{schulz1980critical,maeda2007universal}
The magnetization is $\langle S^z \rangle = m(T,H) = \int
\frac{dk}{2\pi} \bigl[f_+(k) - f_-(k)\bigr]$
and $Y_D(T,H)$ is
\begin{align}
 Y_D(T,H) &= \frac {3Z_2}{2m(T,H)}\int \frac{dk}{2\pi} 
 \frac v{2\omega_0(k)}\bigl[2f_0(k)
 - f_+(k) - f_-(k)\bigr].
 \label{eq.YD_fermion}
\end{align}
This reduces to Eq.~\eqref{eq.YD_dilute} in the limit
$H, T \to 0$.
We emphasize that there is no free parameter in our theory 
since the renormalization factor $Z_{2}$ in the overall coefficient
of~\eqref{eq.YD_fermion}
has been already determined in~\eqref{eq.Z2}.
As shown in Fig.~\ref{fig.esr}, the free-fermion
approximation \eqref{eq.YD_fermion} explains
the extremum of the ESR shift observed numerically
around the critical field $H = \Delta_0$.

Figure~\ref{fig.ndmap} shows the ESR shift
observed experimentally in
$\mathrm{Ni(C_5H_{14}N_2)_2N_3(PF_6)}$,~\cite{kashiwagi2009ndmap}
which possesses the SIA, and the corresponding numerical result
by the QMC method.
Our FFPT~\eqref{eq.YD_fermion}
successfully accounts for the experimental and
numerical results, including the  gradual approach to the paramagnetic
resonance line $\omega = g_e \mu_B H$ in the high field region.
A dtailed analysis of the ESR shift in the whole range of $H$
will be given in a subsequent publication.~\cite{furuya_shift}


We thank Seiichiro Suga for giving us the motivation for this study.
This work is partly supported by
Grant-in-Aid for Scientific Research No. 21540381 (M.O.),
the Global COE Program ``The Physical Sciences Frontier'' (S.C.F.),
both from MEXT, Japan, and Grant-in-Aid from JSPS  (Grant
No.09J08714) (S.T.).
M.O. also acknowledges the Aspen Center for Physics
where a part of this work was carried out
(supported by U.S. NSF Grant No. 1066293). 
We thank the ALPS project for providing the QMC code.
Numerical calculations were performed at the
ISSP Supercomputer Center of the University of Tokyo.


%

\end{document}